\documentclass[runningheads,a4paper]{llncs}

\usepackage{amsfonts}
\usepackage{dsfont}

\usepackage{mathtools}

\usepackage{amssymb}

\usepackage{amsmath}

\usepackage{amsthm}

\usepackage{graphicx}

\usepackage{epstopdf}

\usepackage[hidelinks]{hyperref}
\usepackage{subfig}
\usepackage{caption}

\usepackage{float}

\makeatletter

\def\holdocspecials{\do\ \do\$\do\&%
  \do\#\do\^\do\^^K\do\_\do\^^A\do\%}

\def\holtt{\trivlist \item[]\if@minipage\else\vskip\parskip\fi
\leftskip\@totalleftmargin\rightskip\z@
\parindent\z@\parfillskip\@flushglue\parskip\z@
\@tempswafalse \def\par{\if@tempswa\hbox{}\fi\@tempswatrue\@@par}
\obeylines \tt \let\do\@makeother \holdocspecials
 \frenchspacing\@vobeyspaces}

\makeatother

\newlength{\hsbw}
\newcommand\HOLSpacing{13pt}

   \newcommand\hilbert{\varepsilon}

   \newcommand{\Cond}{\(\Longrightarrow\)}
   \newcommand{\Eqv}{\(\equiv\)}
   \newcommand{\Iff}{\(\Leftrightarrow\)}
   \newcommand{\Fa}{\(\scriptsize \forall\)}
   \newcommand{\Et}{\(\exists\)}
   \newcommand{\Eu}{\(\exists_{unique}\)}

   \newcommand{\Func}{\(\to\)}
   \renewcommand{\Bar}{\(\mid\)}

   \newcommand{\Lam}{\(\lambda\)}
   \newcommand{\Plus}{\(+\)}
   \newcommand{\Minus}{\(-\)}
   \newcommand{\Prime}{\('\)}
   \newcommand{\Und}{\_}
   \newcommand{\Lt}{\(<\)}
   \newcommand{\Gt}{\(>\)}
   \newcommand{\Leq}{\(\leq\)}
   \newcommand{\Geq}{\(\geq\)}
   \newcommand{\Eq}{\(=\)}

\newcommand{\Hilbert}{\(\hilbert\)}
\newcommand{\Imp}{\(\Rightarrow\)}
\newcommand{\Conj}{\(\wedge\)}
\newcommand{\Disj}{\(\vee\)}
\newcommand{\Neg}{\(\neg\)}
\newcommand{\Pnd}{\(\Diamond\)}

\long\def\rechol#1#2#3{\let\next=\rechol\def\postnext{#2#3}\ifx#1\end
\let\next=\relax\def\postnext{\relax}
\else\ifx#1!\Fa                                          
\else\ifx#1@\Hilbert                                     
\else\ifx#1\#\Pnd                                        
\else\ifx#1'\Prime                                       
\else\ifx#1~\Neg                                         
\else\ifx#1\~\Neg
\else\ifx#1_\Und                                         
\else\ifx#1+\Plus
\else\ifx#1\/\Disj                                       
\else\ifx#1\.\Lam                                        
\else\ifx#1>\ifx#2=\Geq\def\postnext{#3}\else\Gt\fi      
\else\ifx#1?\ifx#2!\Eu\def\postnext{#3}\else\Et\fi       
\else\ifx#1-\ifx#2\>\Func\def\postnext{#3}               
	    \else\Minus\fi				 
\else\ifx#1|\ifx#2-\Turns\def\postnext{#3}\else\Bar\fi   
\else\ifx#1<\ifx#2=\ifx#3>\Iff\def\postnext{}            
                   \else\Leq\def\postnext{#3}\fi         
            \else\Lt\fi
\else\ifx#1=\ifx#2=\ifx#3>\Imp\def\postnext{}            
                   \else\Eqv\def\postnext{#3}\fi         
            \else\ifx#2>\Cond\def\postnext{#3}
                 \else\Eq\fi\fi
\else\ifx#1/\ifx#2\^^M\Conj\par\def\postnext{#3}         
            \else\ifx#2\ \Conj\ \def\postnext{#3}\else#1\fi\fi  
\else#1\fi\fi\fi\fi\fi\fi\fi\fi\fi\fi\fi\fi\fi\fi\fi\fi\fi\fi
\expandafter\next\postnext}

\usepackage{epsfig}

\usepackage{array,longtable}

\newcommand{\kw}[1]{{\color{blue}\textsf#1}}

\newcommand{\D}{\mathcal{D}}
\newcommand{\holl}{\textsf{HOL Light}}
 \usepackage{hyperref}
\usepackage{mdframed}

\usepackage{multirow}

\usepackage{url}

\PassOptionsToPackage{hyphens}{url}\usepackage{hyperref}

\urldef{\mailsa}\path|{adnan.rashid, osman.hasan}@seecs.nust.edu.pk|
\urldef{\mailsb}\path|siddiu3@mcmaster.ca|

\newcommand{\keywords}[1]{\par\addvspace\baselineskip
\noindent\keywordname\enspace\ignorespaces#1}

\begin{document}

\mainmatter  

\title{Formal Verification of Platoon Control Strategies}

\titlerunning{Formal Verification of  Platoon Control Strategies}

\author{Adnan Rashid${}^1$, Umair Siddique${}^2$ \and Osman Hasan${}^1$}

%
\authorrunning{A. Rashid, U. Siddique and O. Hasan}


\institute{${}^1$School of Electrical Engineering and Computer Science (SEECS)\\
National University of Sciences and Technology (NUST)\\
Islamabad, Pakistan\\
\mailsa\\
${}^2$Department of Computing and Software \\
McMaster University, Hamilton, Canada\\
\mailsb\\
}
%
%
\maketitle

\begin{abstract}
Recent developments in autonomous driving, vehicle-to-vehicle communication and smart traffic controllers have provided a hope to realize platoon formation of vehicles. The main benefits of vehicle platooning include improved safety, enhanced highway utility, efficient fuel consumption and reduced highway accidents. One of the central components of reliable and efficient platoon formation is the underlying control strategies, e.g., constant spacing, variable spacing and dynamic headway. In this paper, we provide a generic formalization of platoon control strategies in higher-order logic. In particular, we formally verify the stability constraints of various strategies using the libraries of multivariate calculus and Laplace transform within the sound core of \holl~proof assistant. We also illustrate the use of verified stability theorems to develop runtime monitors for each controller, which can be used to automatically detect the violation of stability constraints in a runtime execution or a logged trace of the platoon controller. Our proposed formalization has two main advantages: 1) it provides a framework to combine both static (theorem proving) and dynamic (runtime) verification approaches for platoon controllers; and 2) it is inline with the industrial standards, which explicitly recommend the use of formal methods for functional-safety, e.g., automotive ISO $26262$.
\end{abstract}


\section{Introduction}
Autonomous cars seem to be just around the corner, as most of the car manufacturers (e.g., Tesla, BMW, Toyata, Nissan, Ford, Jaguar Land Rover, etc.) and even silicon valley players (e.g., Intel and Nvidia) claim that fully autonomous vehicles will be on the road around $2020$~\cite{autonomous_cars,autonomous_MIT}. Such a speedy development in autonomous driving is motivated by the fact that the autonomous cars will be more safe and crashless than the human driven cars. For example, the human error is to blame for up to $90\%$ of the $1.2$ million deaths that occur each year from car accidents around the world~\cite{human_error}. Like various fields of science and engineering, the developments in autonomous driving have opened the doors to many other interesting fields, for example, \emph{automated vehicle platooning} is one of the most benefitting fields.

\begin{figure}[ht!]
	\centering
\scalebox{0.255}
{\includegraphics[trim={5.0 0.4cm 5.0 0.4cm},clip]{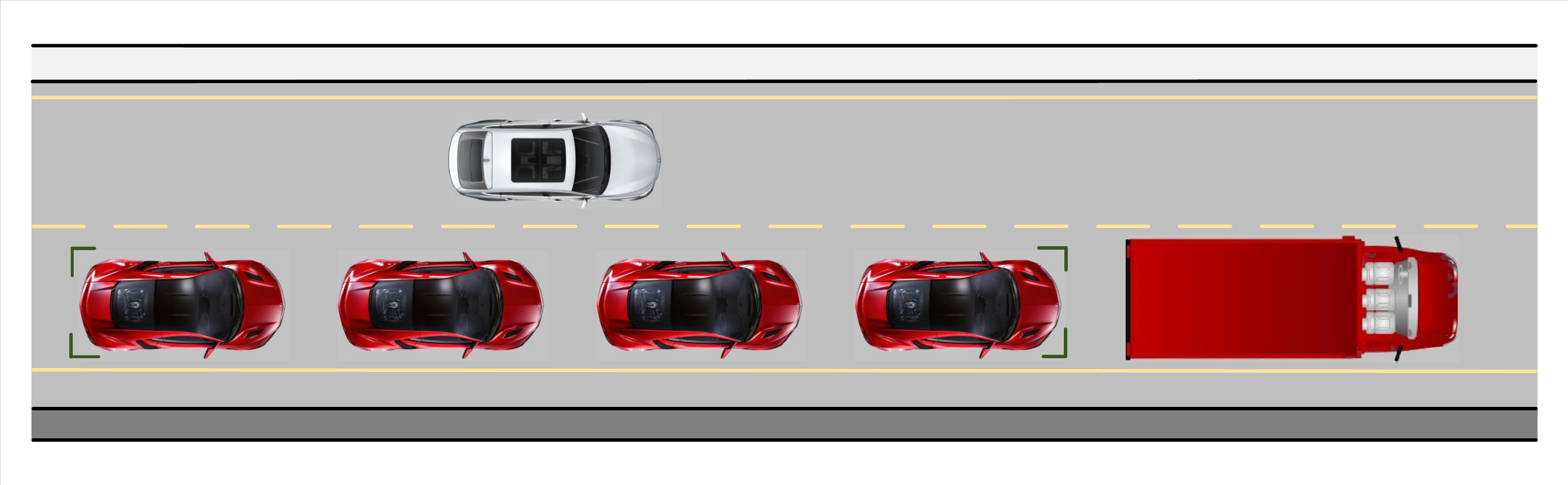}}
	\caption{Platoon of Vehicles}\label{Fig:Platoon}
	\label{FIG:platoon}
\end{figure}

A \emph{platoon} is a group of vehicles (as shown in Figure~\ref{FIG:platoon}) that travels in a close proximity to one another, nose-to-tail, at highway speeds. Vehicle platoons have

\noindent been proposed since at least the early $1980$'s even before we had wireless communication, global positioning system (GPS) and commercially available radar sensors. However, given the exceptional capabilities and reliability of the autonomous cars, vehicle platooning can become a reality using a mix of available technologies such as drive-by-wire steering~\cite{changfu2006development}, radar cruise control~\cite{van2006impact} and lane keep assist systems~\cite{kawazoe2002lane} to name a few. Some of the main advantages of the vehicle platooning are increased road capacity, reduction in drag and improved fuel economy, improved traffic congestion strategies~\cite{fernandes2012platooning} and reduced roadside accidents due to the autonomous features, e.g., collision detection~\cite{biswas2006vehicle} and automatic emergency braking~\cite{yi2000adaptive}.

The stability of the automated vehicles in a platoon, individually or as a group, depends on the interaction of the vehicles and is vital for an uninterrupted flow of traffic and a better throughput. A \emph{stable platoon} ensures that the vehicles should not collide with each other while maintaining a safe inter-vehicle spacing bound. In practice, the stability of platoon is ensured by two types of controllers, i.e., autonomous and non-autonomous~\cite{eyre1998simplified}. The autonomous controllers use the on-board sensors for determining the speed and position of the connected vehicles, whereas the non-autonomous controllers are based on other forms of the inter-vehicle communication. Furthermore, communication amongst controllers is either unidirectional or bidirectional, based on the information shared between the neighbouring vehicles. Similarly, various strategies can be used for the platoon control, such as constant spacing, variable spacing and variable time headway.

Traditionally, the platoon controllers are analyzed using informal approaches including paper-and-pencil based proofs~\cite{eyre1998simplified} and numerical simulations~\cite{barooah2009mistuning}. These informal approaches have known limitations when used in safety-critical domains, for example, missing assumptions and even wrong derivations in hand-driven proofs, and inherent incompleteness of the numerical algorithms, respectively. Considering these facts, it is natural to think about complementing traditional analysis approaches with formal methods for developing reliable  platoon controllers. In this direction, model checking has been used to verify the high-level models of the platoon controllers using the temporal logic based properties~\cite{kamali2017formal,dolginova1998safety,wongpiromsarn2008formal}. In all these approaches, the authors considered the vehicles platoon and their controllers as a discrete-time system by modeling them as automata and verified their properties, such as safety and inter-vehicle spacing bound properties.
Thus, these model checking based analysis lacks the physical analysis of the platoon, which requires modeling and reasoning of control strategies using differential equations and their frequency domain stability analysis using Laplace transform.
Similarly, Mashkoor et al.~\cite{mashkoor2012formal} used higher-order logic to formally reason about the cyber-physical transportation system. The authors used random variables to model the unpredictable elements of the system and formally conducted a probabilistic analysis of the transportation system without considering the dynamic aspects of the system.
In this paper, we propose a higher-order logic based framework to formally model and verify the stability of various types of platoon controllers using the \holl~proof assistant~\cite{harrison1996hol}. Consequently, we utilize the verified results to construct monitors, which ensure the platoon stability at runtime. The main reasons for using higher-order logic and \holl~include the expressibility to represent the platoon controllers, which are modeled using differential equations in time-domain and the Laplace transform in frequency-domain. Moreover, \holl~has the smallest trusted core (i.e., approximately $400$ lines of Ocaml code) amongst all  other HOL proof assistants and the underlying logic kernel has been verified in the CakeML project~\cite{kumar2016self}.

The main contributions of the paper are as follows:

\begin{itemize}
	\item[$\bullet$] Deep embedding based formalization of platoon controller types, configurations and strategies along with the associated differential equations based functional models.
	\item[$\bullet$] Formal derivation of the Laplace domain transfer functions using the formalized libraries of multivariate calculus~\cite{harrison2013hol} and the Laplace transform~\cite{taqdees2013formalization,rashid2017tmformalization} in the \holl~proof assistant.
	\item[$\bullet$] Formal verification of the platoon control strategies based on the formal models of various controllers.
	\item[$\bullet$] Development of the stability monitors for each type of the controllers and demonstration of their violation detection capability on a pseudo-randomly generated traces of a platoon controller.
\end{itemize}

The source code of our \holl~development is available for download at~\cite{adnan2018fvpcs} and thus can be used by the other researchers and engineers interested in the design and verification of the platoon controllers.

The rest of the paper is organized as follows: Section~\ref{SEC:prelim} presents an overview of the \holl~proof assistant along with the formalization of the Laplace transform. Section~\ref{SEC:for_mod_platoon_cv} provides the formal modeling of the platoon controller and its stability. We provide the formal verification of the platoon control strategies and the stability constraints in Section~\ref{SEC:form_ver_platoon_control_strag}. Section~\ref{SEC:application} describes the construction of the stability monitors. Finally, Section~\ref{SEC:conc} concludes the paper and highlights some future research directions.


\section{Preliminaries} \label{SEC:prelim}
This section presents a brief introduction to the \holl~proof assistant and its multivariate calculus and the Laplace transform theories, which are extensively used in the rest of the paper.

\subsection{Theorem Proving and \holl~Proof Assistant}\label{multi_cal_theories}

Theorem proving is a widely adapted formal verification technique, which is concerned with constructing the proofs of the mathematical theorems using a computer program (called \emph{theorem prover or proof assistant})~\cite{harrison_book}.
Theorem proving systems have been commonly used for verifying the properties of the software and hardware systems. For example, a hardware designer can certify a digital circuit by modeling its behavior by some predicates and verifying its different properties using Boolean algebra. Similarly, a mathematician can verify the transitive property of the ordering of real numbers using some basic axioms of real numbers theory. These properties are expressed as theorems using some logic, such as propositional, first-order or higher-order logic, based on the required expressiveness. For example, using the higher-order logic is advantageous over the first-order logic as it provides the additional quantifiers and is more expressive as well. Moreover, higher-order logic can better describe the complex mathematical concepts including multivariate calculus, transcendental functions and topological spaces. Once such a mathematical theory is developed inside a proof assistant, we say that it is formalized.

\holl~\cite{harrison1996hol} is an  interactive theorem proving environment for constructing the mathematical proofs. The main implementation of \holl~is done in a functional programming language, Objective CAML (OCaml), which is originally developed to automate the mathematical proofs~\cite{Ocaml_ref}. The logical kernel of \holl~is of approximately $400$ lines of OCaml code and its main components are its types, terms, theorems, rules of inference, and axioms.  A theory in \holl~consists of types, constants, definitions, axioms and theorems. The \holl~theories are ordered in a hierarchical fashion and the child theories can inherit the types, definitions and theorems of the parent theories. Every new theorem has to be verified based on the primitive inference rules and basic axioms or already verified theorems present in~\holl, which ensures the soundness of this technique.

\subsection{Multivariable Calculus and Laplace Transform Theories}

\holl~provides an extensive support for the analysis of physical systems based on multivariate calculus theories, which include derivatives, integration, transcendental theory, topology, vector analysis and Laplace transform theory. Table~\ref{TAB:lap_trans_for} presents some definitions from the Laplace transform theory of~\holl, which includes the Laplace transform, Laplace existence and the exponential-order conditions, and the differential equation of order $n$. Interested readers can refer to~\cite{taqdees2013formalization,rashid2017tmformalization} for more details about this theory. It is extensively used in our proposed verification of the platoon control strategies for the automated vehicles.

\vspace*{-0.5cm}


\begin{table}[!ht]
	\caption{Laplace Transform}
	\label{TAB:lap_trans_for}
	\begin{tabular}{ |p{3.7cm}|p{8.2cm}| }
		\hline 
		Mathematical Form & Formalized Form   \\  \hline \hline
		\multicolumn{2}{|c|}{\textbf{Laplace Transform}} \\ \hline
		{ {$\begin{array} {lcl} \mathcal{L} [f(t)] = F(s) =  \\
				\int_{0}^{\infty} {f(t)e^{-s t}} dt, \ s \ \epsilon \  \mathds{C}
				\end{array}$}  }
		&
		{$\begin{array} {lcl} \textup{\texttt{\textsf{ $\vdash$ $\forall$ s f. \textbf{laplace\_transform} f s =   } }} \\
			\textup{\texttt{ \textsf{  \hspace*{0.5cm}  integral \{t $|$ \&0 $<=$ drop t\} } }}  \\
			\textup{\texttt{ \textsf{  \hspace*{1.2cm}  ($\lambda$t. cexp ($--$(s $\ast$ Cx (drop t))) $\ast$ f t) } }}
			\end{array}$}  \\ \hline
		\multicolumn{2}{|c|}{\textbf{Laplace Existence}} \\ \hline
		\vspace*{-0.6cm}
		$f$ is piecewise smooth and is of exponential order on the positive real line
		&
		{$\begin{array} {lcl} \textup{\texttt{ \textsf{ $\vdash$ $\forall$ s f. \textbf{laplace\_exists} f s $\Leftrightarrow$ } }} \\
			\textup{\texttt{\textsf{  \hspace*{0.2cm} ($\forall$ b. f piecewise\_differentiable\_on }  }}  \\
			\textup{\texttt{\textsf{  \hspace*{3.0cm}  interval [lift (\&0),lift b] ) $\wedge$  } }}  \\
			\textup{\texttt{\textsf{  \hspace*{0.2cm}  ($\exists$ M a. Re s $>$ drop a $\wedge$ exp\_order\_cond f M a) } }}
			\end{array}$}  \\ \hline
		\multicolumn{2}{|c|}{\textbf{Exponential-order Condition}} \\ \hline
		\vspace*{-0.6cm}
		There exist a constant $a$ and a positive constant $M$ such that $|f (t)| \leq Me^{at}$
		&
		{$\begin{array} {lcl} \textup{\texttt{ \textsf{ $\vdash$ $\forall$ f M a. \textbf{exp\_order\_cond} f M a $\Leftrightarrow$ \&0 $<$ M $\wedge$  } }} \\
			\textup{\texttt{ \hspace*{0.2cm}  ($\forall$ t. \&0 $<=$ t $\Rightarrow$    }}  \\
			\textup{\texttt{\textsf{  \hspace*{0.5cm}  norm (f (lift t)) $<=$ M $\ast$ exp (drop a $\ast$ t)) } }}
			\end{array}$}  \\ \hline
		\multicolumn{2}{|c|}{\textbf{Differential Equation of Order $n$}} \\ \hline
		{ {$\begin{array} {lcl} \textit{Differential} \ \textit{Equation} & =   \\
				\sum\limits_{k = 0}^{n} {{\alpha}_k \dfrac{d^kf}{{dt}^k}} \end{array}$}  }&
		{$\begin{array} {lcl} \textup{\texttt{\textsf{$\vdash$ $\forall$ n f t. \textbf{diff\_eq\_n\_order} n lst f t =    }  }} \\
			\textup{\texttt{\textsf{ \hspace*{0.5cm} vsum (0..n) ($\lambda$k. EL k [$\alpha_1,\alpha_2,...,\alpha_k$] $\ast$ }}}  \\
			\textup{\texttt{\textsf{  \hspace*{1.5cm} higher\_order\_derivative k f t) }}}
			\end{array}$}  \\ \hline
	\end{tabular}
\end{table}



\section{Formal Modeling of Platoon Controller and Stability}\label{SEC:for_mod_platoon_cv}
In this section, we present the formal modeling of a platoon controller based on its types, configurations and the underlying strategies along with the concept of the platoon stability.

\subsection{Formalization of Platoon Controller}\label{SUBSEC:for_mod_platoon}
The connected vehicles in a platoon are widely characterized by the controllers, which are mainly responsible for their automated operation. The platoon controllers are generally of two types namely autonomous and non-autonomous.
\begin{itemize}
  \item[$\bullet$]  \emph{Autonomous controllers} use the on-board sensors for determining the speed and position of the connected vehicles.
  \item[$\bullet$]  \emph{Non-autonomous controllers} are based on some other form of the inter-vehicle communication.
\end{itemize}

The information sharing among the neighbouring vehicles is either unidirectional or bidirectional depending upon the configuration of the platoon controllers.

\begin{itemize}
  \item[$\bullet$] \emph{Unidirectional configuration} allows a controller to use the information about the relative distance and velocity of only the preceding vehicle.
  \item[$\bullet$] \emph{Bidirectional controller} can access the information about the relative distance and velocity of both the proceeding and preceding vehicles by considering their individual masses.
\end{itemize}

The autonomous controllers can adapt different strategies to maintain the stable operation of the platoon along the highway. In general, controllers utilize three strategies namely constant spacing, variable spacing and variable time-headway.
\begin{itemize}
  \item[$\bullet$] \emph{Constant spacing policy} requires that each vehicle maintains a constant distance (spacing) with its preceding vehicle in a platoon.
  \item[$\bullet$] \emph{Variable spacing policy} allows a variable inter-vehicle spacing, which depends on the velocity of the vehicles in a platoon. For example, a faster moving vehicle creates more inter-vehicle space between itself and its proceeding vehicle. It is also known as the \emph{constant time headway spacing}.
  \item[$\bullet$] \emph{Variable time headway} policy imposes constraints on the relative velocity rather than the absolute velocity of the vehicle in contrast to the constant time headway spacing policy, which results into large inter-vehicle spaces and thus decreases the throughput of the highway traffic.
\end{itemize}

In our formalization, we model the types of the controller, its configurations and strategies as enumerated datatype using the built-in \texttt{\textsf{define\_type}} mechanism in \holl.

\begin{mdframed}
{
\small
\textup{\texttt{
\hspace*{-0.36cm}  \kw{type} \textsf{\textbf{controller\_type} = autonomous} $\mid$ \textsf{non\_autonomous} \\
\kw{type} \textsf{\textbf{configuration} = unidirectional $\mid$ bidirectional} \\
\kw{type} \textsf{\textbf{strategy} = constant\_spacing $\mid$ variable\_spacing $\mid$ var\_time\_headway}
}}}
\end{mdframed}

We model a platoon as a tuple $(x,n,m,k,c,ch,vd,h0,ca,cd)$, where the description and the type of each parameter is given in Table~\ref{TAB:platoon_parameters}. Indeed, these parameters
characterize various physical aspects of the vehicles in a platoon (e.g., the horizontal distance $x$, the number of vehicles in a platoon $n$ and the mass of a vehicle $m$). In \holl, we
formalize the platoon tuple $(x,n,m,k,c,ch,vd,h0,ca,cd)$  and controller tuple $(controller\_type,configuration,strategy,platoon)$ as type abbreviations:

\begin{mdframed}
{
\small
\textup{\texttt{
\hspace*{-0.4cm}  \kw{type\_abbrev} \textsf{\textbf{platoon}:(x $\times$ n $\times$ m $\times$ k $\times$ c $\times$ h $\times$ ch $\times$ vd $\times$ h0 $\times$ ca $\times$ cd)} \\
\hspace*{-0.2cm} \kw{type\_abbrev} \textsf{\textbf{controller}:(controller\_type $\times$ configuration $\times$ strategy $\times$ platoon)}
}}}
\end{mdframed}


\noindent It is important to note that \texttt{platoon} contains a unique mass $m$, which implies that we only consider a platoon with identical vehicles as shown in Figure \ref{Fig:Platoon}.

\begin{table}[!ht]
\caption{Data Types for Platoon Parameters}
\label{TAB:platoon_parameters}
\begin{center}
	\begin{tabular}{ |c|c| }
		\hline \hline
	Parameter Description & Type   \\ 	\hline \hline
	    Horizontal distance & \textsf{x:$\mathbb{N}\rightarrow(\mathbb{R} \rightarrow \mathbb{C})$}  \\
	   Number of vehicles & \textsf{n:$\mathbb{N}$} \\
	  Mass of a vehicle & \textsf{m:$\mathbb{R}$} \\
	   Disturbance constant  &  \textsf{k:$\mathbb{R}$}\\
	 Fluctuations  constant &  \textsf{c:$\mathbb{R}$}\\
	 Time headway  &  \textsf{h:$\mathbb{R}$}\\
	Fluctuations due to time headway  & \textsf{ch:$\mathbb{R}$} \\
	 Desired platoon speed &  \textsf{vd:$\mathbb{R}$}\\
	 Nominal value of time headway   & \textsf{h0:$\mathbb{R}$}\\
   Additional fluctuations with respect to platoon leader &  \textsf{ca:$\mathbb{R}$}\\
   Additional fluctuations with respect to ``virtual" mass &  \textsf{cd:$\mathbb{R}$}\\
	   	\hline
	\end{tabular}
\end{center}
\end{table}


In order to ensure that the given parameters of a platoon indeed represent a valid platoon, we formalize the associated constraints as a predicate \texttt{\textsf{is\_valid\_platoon}} (Definition \ref{DEF:valid_platoon}).
For example, the mass $m$ should always be greater than $0$ and the number of vehicles in a platoon should be greater than $1$.

\begin{mdframed}
\begin{definition}
\label{DEF:valid_platoon}
\emph{Valid Platoon} \\{
\small
\textup{\texttt{
$\vdash$ \textsf{\textbf{is\_valid\_platoon} (x,n,m,k,c,h,ch,vd,h0,ca,cd)} $\Leftrightarrow$ \textsf{0 $<$ m $\wedge$ 0 $<$ k $\wedge$ 0 $<$ c $\wedge$} \\
\hspace*{2.0cm}   \textsf{0 $<$ h $\wedge$ 0 $<$ ch $\wedge$ 0 $<$ vd $\wedge$ 0 $<$ h0 $\wedge$ 0 $<$ ca $\wedge$ 0 $<$ cd $\wedge$ 1 $<$ n}
}}}
\end{definition}
\end{mdframed}

\subsection{Formalization of the Platoon Stability}\label{SUBSEC:for_mod_platoon_stability}

The \emph{stability} is an important property of a vehicle platoon, which describes the capability of a platoon to attenuate the oscillations introduced by the leader or any other vehicle in the platoon. In general, such oscillations can be considered in terms of various signals. e.g., the position error between the vehicles or the relative acceleration of connected vehicles. In this paper, we consider the notion of stability with respect to the position error between the vehicles. Formally, a platoon is stable if any oscillation with respect to the position error diminishes out as it propagates towards the tail of the platoon.  The platoon stability in longitudinal direction is mathematically expressed as a norm condition on spacing errors in the frequency domain, as given in the following equation:

\begin{equation}\label{EQ:platoon_stability}
\bigg| \bigg|  \frac{z_n (i \omega)}{z_{n - 1} (i \omega)} \bigg| \bigg| < 1, \ \ \ n = 2, 3, 4, ...
\end{equation}

\noindent where $z_{n - 1}$ is the spacing error between the vehicle $n - 1$ and its proceeding vehicle $n$, i.e., it is the deviation from the desired inter-vehicle spacing for vehicle $n - 1$. If $x_{n - 1}$ is the inter-vehicle spacing between the vehicle $n - 1$ and its preceding vehicle $n - 2$ and $x_n$ is the inter-vehicle spacing between the vehicle $n$ and its preceding vehicle $n - 1$, then the spacing error between vehicles $n - 1$ and $n$ is given by $z_{n - 1} = x_{n - 1} - x_n$.
Similarly, $z_n$ represents the spacing error between the vehicle $n$ and its proceeding vehicle $n + 1$, i.e., $z_n = x_n - x_{n + 1}$. In case of all the desired inter-vehicle spacings are same, i.e., $x_n = x_{n - 1} = ... = x_1$, then this leads to the zero spacing errors, i.e., $z_n = z_{n - 1} = ... = z_1 = 0$.
We formalize platoon stability in \holl~as follows:

\begin{mdframed}
\begin{definition}
\label{DEF:is_stable_platoon}
\emph{Stable Platoon} \\
{
\small
\textup{\texttt{ \vspace*{0.2cm}
$\vdash$ \textsf{$\forall$ s x y. \textbf{transfer\_function} s x y} $\Leftrightarrow$ $\mathsf{\dfrac{laplace\_transform \ y \ s}{laplace\_transform \ x \ s}}$
}}}  \\
{
\small
\textup{\texttt{ \vspace*{0.1cm}
$\vdash$ \textsf{$\forall$ $\omega$ x y. \textbf{frequency\_response} $\omega$ x y} $\Leftrightarrow$ \textsf{transfer\_function (i$\omega$) x y}
}}}  \\
{
\small
\textup{\texttt{ \vspace*{0.1cm}
$\vdash$ \textsf{$\forall$ $\omega$ z. \textbf{is\_stable\_platoon} $\omega$ z n} $\Leftrightarrow$ \vspace*{0.1cm}  \\
\hspace*{2.0cm}  \textsf{\bigg |\bigg | frequency\_response $\omega$ ($\lambda$t. z (n)) ($\lambda$t. z (n - 1)) \bigg |\bigg | $<$ 1}
}}}
\end{definition}
\end{mdframed}

\noindent where the predicate \texttt{\textsf{is\_stable\_platoon}} accepts the parameters \texttt{\textsf{z}}:$\mathbb{N}\rightarrow(\mathbb{R} \rightarrow \mathbb{C})$, which represents the complex-domain representation of the spacing error, angular frequency $\omega$:$\mathbb{R}$ and number of vehicles \texttt{\textsf{n}}, and returns the condition that the complex norm of the transfer function at $s = i \omega$, i.e., $\frac{Z_n(i \omega)}{Z_{n - 1}(i \omega)}$ is always less than $1$ for every vehicle in the platoon.

This concludes our fundamental formalization of the platoon controller and the corresponding stability. We build upon the concepts, formalized in this section, to formalize various control strategies and verify their correctness
in the next section.

\section{Formal Verification of the Platoon Control Strategies}\label{SEC:form_ver_platoon_control_strag}

In this section, we first present the formalization of an autonomous unidirectional controller with constant spacing policy. Indeed, the main intention is to demonstrate the formalization steps, i.e., formal modeling of the controller dynamics in higher-order logic, formalization of the necessary constraints, and the formal verification of the stability theorem. Building upon these steps, we next present its generalization to all types of controllers along with the verification of a generalized stability theorem.

\subsection{Autonomous Unidirectional Controller}\label{SUBSEC:autonom_unidirec_control}

Generally, the dynamics of platoon controllers are characterized by a set of differential equations, which interrelate the parameters of the platoon. The schematic representation of the platoon of vehicles having autonomous unidirectional controller with constant spacing policy is depicted in Figure~\ref{FIG:Platoon_aut_uni_cont_cs}. It consists of $n$ interconnected vehicles of identical masses, i.e., $m_1 = m_2 = ... = m_{n - 1} = m_n = m$. The parameters $k$ and $c$ are the disturbance and fluctuation constants, representing the control gains on the relative position and velocity, respectively. Similarly, the parameter $u$ represents the force required by the first vehicle to move forward in the platoon. The mathematical representation of this platoon controller's dynamics are given as the following equation set:

\begin{figure}[!ht]
\centering
\scalebox{0.35}
{
\includegraphics[trim={5.0 0.4cm 5.0 0.4cm},clip]{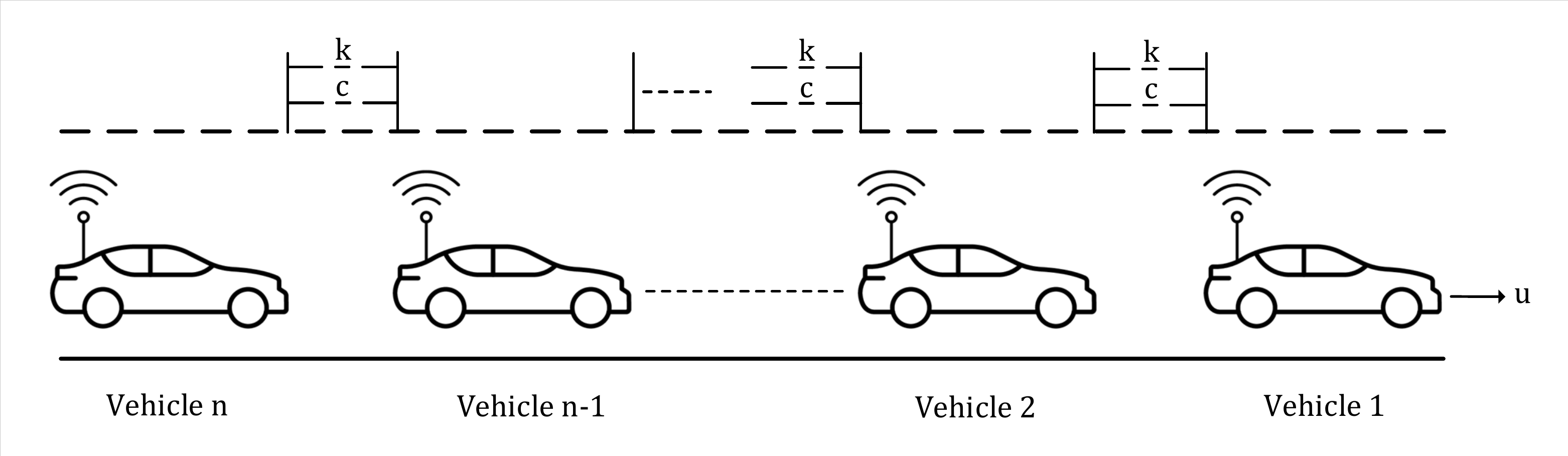}}
\caption{Autonomous Unidirectional Controller with Constant Spacing}
\label{FIG:Platoon_aut_uni_cont_cs}
\end{figure}

{\small
\begin{equation}\label{EQ:diff_eqn_uni_cs}
\begin{split}
\frac{dx_1}{dt} = v_1, \
\frac{dv_1}{dt} = \frac{u}{m}, \
\frac{dx_2}{dt} = v_2  \\
\frac{dv_2}{dt} = \frac{k}{m} x_1 - \frac{k}{m} x_2 + \frac{c}{m} v_1 - \frac{c}{m} v_2  \\
. \\
. \\
\frac{dx_{n-1}}{dt} = v_{n-1}  \\
\frac{dv_{n-1}}{dt} = \frac{k}{m} x_{n-2} - \frac{k}{m} x_{n-1} + \frac{c}{m} v_{n-2} - \frac{c}{m} v_{n-1}  \\
\frac{dx_n}{dt} = v_n  \\
\frac{dv_n}{dt} = \frac{k}{m} x_{n-1} - \frac{k}{m} x_n + \frac{c}{m} v_{n-1} - \frac{c}{m} v_n  \\
\end{split}
\end{equation}
}

\noindent where the variables $x$ and $v$ represent the inter-vehicle spacing and velocity of platoon vehicles, respectively.
Overall, the set of differential equations characterize the dynamics of $n$ vehicles in the platoon depicted in Figure~\ref{FIG:Platoon_aut_uni_cont_cs}. We can rewrite Equation (\ref{EQ:diff_eqn_uni_cs}) in a compact form by eliminating the variable $v$ and representing it in the form of spacing error, i.e., $z$ as:

\begin{equation} \label{EQ:diff_eqn_spac_er_uni_cs}
\frac{d^2z_n}{dt^2} + \frac{c}{m} \frac{dz_n}{dt} + \frac{k}{m} z_n = \frac{c}{m} \frac{dz_{n - 1}}{dt} + \frac{k}{m} z_{n - 1}, \ \ \ n = 2, 3, 4, ...
\end{equation}

We formally model this controller in \holl~as follows:

\begin{mdframed}
\begin{definition}
\label{DEF:platoon_uni_cs}
\emph{Unidirectional Controller with Constant Spacing} \\{
\small
\textup{\texttt{\hspace*{-0.1cm} $\vdash$ $\forall$ k c m ch n vd ca cd h h0 x. \\
\hspace*{0.1cm}  \textsf{\textbf{control\_uni\_cs} (autonomous,unidirectional,constant\_spacing,}  \\
\hspace*{5.0cm}  \textsf{((x,n,m,k,c,h,ch,vd,h0,ca,cd):platoon)) t}  $\Leftrightarrow$  \\
\hspace*{0.40cm}  \kw{let} \textsf{$\mathsf{z_{n-1}}$ =  ($\lambda$t. x (n - 1) t - x (n) t)} \kw{and} \\
\hspace*{0.13cm} \text{\  \  \  \  \ \ } \textsf{$\mathsf{z_{n}}$ = ($\lambda$t. x (n) t - x (n + 1) t) \kw{in}} \\
\hspace*{3.0cm} $\mathsf{\D^2 \ [\frac{k}{m};\frac{c}{m}; 1] \ z_{n}}$ =  $\mathsf{\D^1  \ [\frac{k}{m};\frac{c}{m}] \ z_{n-1}}$
}}}
\end{definition}
\end{mdframed}

\noindent where the operators $\mathsf{\D^1}$ and $\mathsf{\D^2}$ represent the first-order and second-order complex-valued derivatives in \holl, respectively, and thus can be obtained by instantiating $n=1$ and $n=2$ in the predicate \texttt{\textsf{diff\_eq\_n\_order}}, given in Table~\ref{TAB:lap_trans_for}.

We next model some physical constraints associated with the controller model \texttt{\textsf{control\_uni\_cs}}, which include differentiability, existence of the Laplace transform and zero-initial conditions for parameters
$\mathsf{z_{n-1}}$ and $\mathsf{z_{n}}$, as given in Definition \ref{DEF:constraints_uni_cs}.

\begin{mdframed}
\begin{definition}
\label{DEF:constraints_uni_cs}
\emph{Constraints for a Platoon having Autonomous Unidirectional Controller} \\{
\small
\textup{\texttt{
\hspace*{-0.1cm} $\vdash$ $\forall$ \textsf{x n s c m k}. \\
\hspace*{0.5cm}    \textsf{\textbf{constraints\_uni\_cs} x n s c m k} $\Leftrightarrow$    \\
\hspace*{1.1cm}  \kw{let} \textsf{$\mathsf{z_{n-1}}$ =  ($\lambda$t. x (n - 1) t - x (n) t) \kw{and}} \\
\hspace*{1.18cm}   \text{\  \ \ \ } \textsf{$\mathsf{z_{n}}$ = ($\lambda$t. x (n) t - x (n + 1) t) \kw{in}} \\
\hspace*{2.5cm}   \textsf{($\forall$t. differentiable\_higher\_derivative [2,1] [$\mathsf{z_{n-1}}$,$\mathsf{z_{n}}$]) $\wedge$}  \\
\hspace*{2.5cm}   \textsf{\ \ \ \ \ laplace\_exists\_higher\_deriv [2,1] [$\mathsf{z_{n-1}}$,$\mathsf{z_{n}}$] s $\wedge$}  \\
\hspace*{2.5cm}   \textsf{\ \ \ \ \ zero\_initial\_conditions [1,0] [$\mathsf{z_{n-1}}$,$\mathsf{z_{n}}$] $\wedge$}  \\
\hspace*{2.5cm}   \textsf{\ \ \ \ \ non\_zero\_tf\_uni\_cs $\mathsf{z_{n-1}}$ s c m k}
}}}
\end{definition}
\end{mdframed}

\noindent where the first two conjuncts provide the differentiability and the Laplace existence conditions for the second-order and first-order derivatives of the spacing errors  $\mathsf{z_{n-1}}$ and $\mathsf{z_{n}}$, respectively. Similarly,  the next conjunct  imposes the zero-initial conditions for the spacing errors $\mathsf{z_{n-1}}$ and $\mathsf{z_{n}}$, respectively.
Finally, the last conjunct ensures that the transfer function does not include the singularities, i.e., the points at which the denominator of the transfer function becomes infinite or undefined. Mathematically, it is described as \texttt{\textsf{$\mathsf{s^2}$ + $\mathsf{\dfrac{c}{m}}$ s + $\mathsf{\dfrac{k}{m}}$ $\neq$ 0}}.

Our next step is to formally verify that the platoon controller model \texttt{\textsf{control\_uni\_cs}} implies the platoon stability for any number of vehicles. The main purpose of this verification is twofold: 1) identify the stability constraints in terms of the platoon parameters, and 2) utilize verified stability constraints to ensure the stability of a given platoon at any time instant. Indeed this step requires the instantiation of platoon parameters with concrete values (i.e., number of vehicles $n = 10$, mass $m = 1000 kg$, etc.). We verify the following universally quantified stability theorem in \holl.

\begin{mdframed}
\begin{flushleft}
\begin{theorem}
\label{THM:tf_imp_stable_uni_cs}
\emph{Stability of a Platoon having Autonomous Unidirectional Controller} \\{\small
\textup{\texttt{
\textsf{\hspace*{-0.1cm} $\vdash$ $\forall$ k c m ch n vd ca cd h h0 x w. \\
 \hspace*{0.1cm}  \kw{let} s = $\mathsf{\mathit{i} \omega}$ \kw{and}  \\
\hspace*{0.55cm}  p = ((x,n,m,k,c,h,ch,vd,h0,ca,cd):platoon) \kw{and}} \\
\hspace*{0.12cm} \text{\  } \textsf{$\mathsf{z}$ = ($\lambda$ n\ t. x (n) t - x (n + 1) t) \kw{in}} \\
\hspace*{-0.1cm} \text{\  } \textsf{0 $<$ $\omega$ $\wedge$  $\mathtt{\mathsf{\dfrac{2 k}{m} \ < \ {\omega}^2}}$ $\wedge$ }  valid\_platoon p $\wedge$ \textsf{constraints\_uni\_cs x n s c k m $\wedge$ \\
 \hspace*{0.10cm}\text{\  \ } \textsf{$\forall$ t. control\_uni\_cs (autonomous,unidirectional,constant\_spacing,p) t} \\
\hspace*{0.28cm} \text{\  \ } $\Longrightarrow$   is\_stable\_platoon $\omega$ z n}
}}}
\end{theorem}
\end{flushleft}
\end{mdframed}

The main proof of Theorem~\ref{THM:tf_imp_stable_uni_cs} consists of the following steps: 1) rewriting with the Definitions~\ref{DEF:valid_platoon}-\ref{DEF:constraints_uni_cs}, 2) complex arithmetic reasoning and 3) the verification of Lemma 1, which transforms the time-domain model of the platoon controller \texttt{\textsf{control\_uni\_cs}} into its equivalent frequency-domain representation, i.e., transfer function. The verification of Lemma~\ref{LEM:mod_imp_trans_fun_uni_cs} is quite involved due to the reasoning about the Laplace transform in \holl~\cite{taqdees2013formalization}. The formal statement of Lemma~\ref{LEM:mod_imp_trans_fun_uni_cs} is given as follows:

\begin{mdframed}
\begin{flushleft}
\begin{lemma}
\label{LEM:mod_imp_trans_fun_uni_cs}
\emph{Model Implies Transfer Function} \\{\small
\textup{\texttt{
\textsf{\hspace*{-0.1cm} $\vdash$ $\forall$ k c m ch n vd ca cd h h0 x s. \\
\hspace*{0.1cm}  \kw{let} p = ((x,n,m,k,c,h,ch,vd,h0,ca,cd):platoon) \kw{and}} \\
\hspace*{0.12cm} \text{\   } \textsf{$\mathsf{z_{n-1}}$ =  ($\lambda$t. x (n - 1) t - x (n) t) \kw{and}} \\
\hspace*{0.12cm} \text{\  } \textsf{$\mathsf{z_{n}}$ = ($\lambda$t. x (n) t - x (n + 1) t) \kw{in}} \\
\hspace*{0.10cm}\text{\ \  }  valid\_platoon p $\wedge$ \textsf{constraints\_uni\_cs x n s c k m} $\wedge$ \\
 \hspace*{0.10cm}\text{\ \  } \textsf{$\forall$ t. control\_uni\_cs (autonomous,unidirectional,constant\_spacing,p) t}  \\
\hspace*{0.18cm}  \text{\  \ } $\Longrightarrow$ $\mathsf{transfer\_function \ s \ z_{n} \ z_{n-1}}$ = $\mathsf{\dfrac{\dfrac{c}{m}\ s \ + \ \dfrac{k}{m}}{s^2 \ + \ \dfrac{c}{m} \  s + \dfrac{k}{m}}}$
}}}
\end{lemma}
\end{flushleft}
\end{mdframed}

\subsection{Generalized Platoon Controller} \label{SUBSEC:gen_platoon_control}

We formally model various types of platoon control strategies as given in Table~\ref{TAB:for_models_platoon_control_strategies}. We also formalize the physical constraints and verify the stability for each control strategy along the same lines as that of autonomous unidirectional controller presented in Section~\ref{SUBSEC:autonom_unidirec_control}. Finally, we package them in an inductive predicate \textsf{gen\_platoon\_control}, which takes two parameters, i.e.,  \textsf{controller} and time \textsf{t} and returns the predicate describing the physical behavior of the controller. For example, for controller \textsf{(autonomous,unidirectional,constant\_spacing,platoon)}, the inductive predicate \textsf{gen\_platoon\_control} returns~\textsf{control\_uni\_cs}\footnote{We have omitted the formal definition of \textsf{gen\_platoon\_control} for the sake of conciseness, however, interested reader can find the formal definition and \holl~code on the project's webpage~\cite{adnan2018fvpcs}.}.

Finally, we verify a general theorem, which describes the stability constraints for any type of the controller $cc$, as follows:

\begin{mdframed}
\begin{flushleft}
\begin{theorem}
\label{THM:MAIN}
\emph{Stability of a Platoon} \\{\small
\textup{\texttt{
\textsf{\hspace*{-0.1cm} $\vdash$ $\forall$ (cc:controller) s. \\
 \hspace*{0.1cm}  \kw{let} s = $\mathsf{\mathit{i} \omega}$ \kw{and}  \\
 \hspace*{0.55cm}   p = (x,n,m,k,c,h,ch,vd,h0,ca,cd):platoon \kw{and}} \\
 \hspace*{0.50cm}   cc = (ct,cg,sg,p) \kw{and} \\
 \hspace*{0.0cm} \text{\ \  } \textsf{$\mathsf{z}$ = ($\lambda$n \ t. x (n) t - x (n + 1) t) \kw{in}} \\
 \hspace*{0.18cm}\text{\ \  }  \textsf{gen\_stability\_physical\_constraints cc s $\mathsf{\omega}$} $\wedge$ \textsf{$\forall$ t. gen\_platoon\_control cc t} \\
  \hspace*{0.40cm}\text{\  \ } $\Longrightarrow$  is\_stable\_platoon $\omega$ z n}
}}
\end{theorem}
\end{flushleft}
\end{mdframed}

\noindent where the predicate \texttt{\textsf{gen\_stability\_physical\_constraints}} encapsulates the stability and physical constraints of all types of controllers in our formalization. The formal proof of Theorem \ref{THM:MAIN} is based on induction on \texttt{\textsf{cc:controller}} and further induction on the \texttt{\textsf{controller\_type}}, \texttt{\textsf{configuration}} and \texttt{\textsf{strategy}} along with the verified stability theorems for each control strategy (e.g., Theorem~\ref{THM:tf_imp_stable_uni_cs} for autonomous unidirectional controller presented in Section~\ref{SUBSEC:autonom_unidirec_control}).

This  concludes our  formalization of platoon controllers in the \holl~proof assistant. In summary, we formalized the basic notions of the platoon controllers


\begin{table}[H]
\caption{Formal Platoon Models considering Various Control Strategies}
\label{TAB:for_models_platoon_control_strategies}
\begin{tabular}{ |p{8.7cm}|p{13.2cm}| }
		\hline 
  \multicolumn{2}{|l|}{Autonomous Unidirectional Controller with Speed-dependent Spacing \hspace*{2.0cm}} \\ \hline
  \multicolumn{2}{|l|}{ $\begin{array} {lcl} \textup{\texttt{\textsf{  $\vdash$ $\forall$ k c m ch n vd ca cd h h0 x.  } }} \\
\textup{\texttt{  \hspace*{0.1cm}  \textsf{\textbf{control\_uni\_vs} (autonomous,unidirectional,variable\_spacing,}      }}  \\
\textup{\texttt{  \hspace*{5.0cm}  \textsf{((x,n,m,k,c,h,ch,vd,h0,ca,cd):platoon)) t}  $\Leftrightarrow$    }}  \\
\textup{\texttt{  \hspace*{0.39cm} \kw{let}  \textsf{$\mathsf{z_{n-1}}$ =  ($\lambda$t. x (n - 1) t - x (n) t) \kw{and}}  }}  \\
\textup{\texttt{  \hspace*{0.13cm} \text{\  \  \  \  } \textsf{$\mathsf{z_{n}}$ = ($\lambda$t. x (n) t - x (n + 1) t) \kw{in}}  }}  \\
\textup{\texttt{    \hspace*{1.9cm} $\mathsf{\D^2 \ [\frac{k}{m};\frac{c + k \ast h}{m}; 1] \ \mathsf{z_{n}}}$ =  $\mathsf{\D^1  \ [\frac{k}{m};\frac{c}{m}] \ \mathsf{z_{n-1}}}  $   }}
 \end{array}$ }   \\ \hline
  \multicolumn{2}{|l|}{Autonomous Unidirectional Controller with Variable Time Headway } \\ \hline
  \multicolumn{2}{|l|}{ $\begin{array} {lcl} \textup{\texttt{\textsf{  $\vdash$ $\forall$ k c m ch n vd ca cd h h0 x.  } }} \\
\textup{\texttt{  \hspace*{0.1cm}  \textsf{\textbf{control\_uni\_vth} (autonomous,unidirectional,var\_time\_headway,}     }}  \\
\textup{\texttt{  \hspace*{5.0cm}  \textsf{((x,n,m,k,c,h,ch,vd,h0,ca,cd):platoon)) t}  $\Leftrightarrow$    }}  \\
\textup{\texttt{  \hspace*{0.39cm} \kw{let}  \textsf{$\mathsf{z_{n-1}}$ =  ($\lambda$t. x (n - 1) t - x (n) t) \kw{and}}  }}  \\
\textup{\texttt{  \hspace*{0.13cm} \text{\  \  \  \  } \textsf{$\mathsf{z_{n}}$ = ($\lambda$t. x (n) t - x (n + 1) t) \kw{in}}  }}  \\
\textup{\texttt{    \hspace*{1.9cm} $\mathsf{\D^2 \ [\frac{k}{m};\frac{c + k \ast h0 + k \ast ch \ast vd}{m}; 1] \ \mathsf{z_{n}}}$ =  $\mathsf{\D^1  \ [\frac{k}{m};\frac{c + k \ast ch \ast vd}{m}] \ \mathsf{z_{n-1}}}  $   }}
 \end{array}$ }   \\ \hline
  \multicolumn{2}{|l|}{Autonomous Bidirectional Controller with Constant Spacing  } \\ \hline
  \multicolumn{2}{|l|}{ $\begin{array} {lcl} \textup{\texttt{\textsf{  $\vdash$ $\forall$ k c m ch n vd ca cd h h0 x.  } }} \\
\textup{\texttt{  \hspace*{0.1cm}  \textsf{\textbf{control\_bi\_cs} (autonomous,bidirectional,constant\_spacing,}    }}  \\
\textup{\texttt{  \hspace*{5.0cm}  \textsf{((x,n,m,k,c,h,ch,vd,h0,ca,cd):platoon)) t}  $\Leftrightarrow$    }}  \\
\textup{\texttt{  \hspace*{0.39cm} \kw{let}  \textsf{$\mathsf{z_{n-1}}$ =  ($\lambda$t. x (n - 1) t - x (n) t) \kw{and}}  }}  \\
\textup{\texttt{  \hspace*{0.13cm} \text{\  \  \  \  } \textsf{$\mathsf{z_{n}}$ = ($\lambda$t. x (n) t - x (n + 1) t) \kw{in}}  }}  \\
\textup{\texttt{    \hspace*{1.9cm} $\mathsf{\D^2 \ [\frac{2 \ast k}{m};\frac{2 \ast c}{m}; 1] \ \mathsf{z_{n}}}$ =  $\mathsf{\D^1  \ [\frac{k}{m};\frac{c}{m}] \ \mathsf{z_{n-1}}}  $   }}
 \end{array}$ }   \\ \hline
  \multicolumn{2}{|l|}{Autonomous Bidirectional Controller with Speed-dependent Spacing  } \\ \hline
  \multicolumn{2}{|l|}{ $\begin{array} {lcl} \textup{\texttt{\textsf{  $\vdash$ $\forall$ k c m ch n vd ca cd h h0 x.  } }} \\
\textup{\texttt{  \hspace*{0.1cm}  \textsf{\textbf{control\_bi\_vs} (autonomous,bidirectional,variable\_spacing,}    }}  \\
\textup{\texttt{  \hspace*{5.0cm}  \textsf{((x,n,m,k,c,h,ch,vd,h0,ca,cd):platoon)) t}  $\Leftrightarrow$    }}  \\
\textup{\texttt{  \hspace*{0.39cm} \kw{let}  \textsf{$\mathsf{z_{n-1}}$ =  ($\lambda$t. x (n - 1) t - x (n) t) \kw{and}}  }}  \\
\textup{\texttt{  \hspace*{0.13cm} \text{\  \  \  \  } \textsf{$\mathsf{z_{n}}$ = ($\lambda$t. x (n) t - x (n + 1) t) \kw{in}}  }}  \\
\textup{\texttt{    \hspace*{1.9cm} $\mathsf{\D^2 \ [\frac{2 \ast k}{m};\frac{2 \ast c + k \ast h}{m}; 1] \ \mathsf{z_{n}}}$ =  $\mathsf{\D^1  \ [\frac{k}{m};\frac{c}{m}] \ \mathsf{z_{n-1}}}  $   }}
 \end{array}$ }   \\ \hline
\multicolumn{2}{|l|}{Non-autonomous Controller considering Communication of Leader's Current Velocity  } \\ \hline
  \multicolumn{2}{|l|}{ $\begin{array} {lcl} \textup{\texttt{\textsf{  $\vdash$ $\forall$ k c m ch n vd ca cd h h0 x.  } }} \\
\textup{\texttt{  \hspace*{0.1cm}  \textsf{\textbf{control\_clcv} non\_autonomous ((x,n,m,k,c,h,ch,vd,h0,ca,cd):platoon) t}  $\Leftrightarrow$    }}  \\
\textup{\texttt{  \hspace*{0.39cm} \kw{let}  \textsf{$\mathsf{z_{n-1}}$ =  ($\lambda$t. x (n - 1) t - x (n) t) \kw{and}}  }}  \\
\textup{\texttt{  \hspace*{0.13cm} \text{\  \  \  \  } \textsf{$\mathsf{z_{n}}$ = ($\lambda$t. x (n) t - x (n + 1) t) \kw{in}}  }}  \\
\textup{\texttt{    \hspace*{1.9cm} $\mathsf{\D^2 \ [\frac{k}{m};\frac{c + ca}{m}; 1] \ \mathsf{z_{n}}}$ =  $\mathsf{\D^1  \ [\frac{k}{m};\frac{c}{m}] \ \mathsf{z_{n-1}}}  $   }}
 \end{array}$ }   \\ \hline
	\end{tabular}
\end{table}

\noindent using the new type definition and corresponding physical and stability constraints. The notable feature of our formalization is its generic nature, as we can model a platoon controller with any number of vehicles composed of basic controller types, configurations and strategies. Moreover, the physical and stability constraints are explicitly present in our formally verified stability theorems, which may get ignored in the conventional platoon analysis and may result into an unstable platoon interrupting the traffic flow on the highways.
In the next section, we describe the utilization of our verified results in \holl~to develop stability monitors for automatically detecting the violations of the stability constraints.


\section{From Verified Controller to Stability Monitors} \label{SEC:application}

\vspace*{-0.3cm}

Static formal verification approaches, such as theorem proving, provide an effective way to formally model and verify digital hardware, its underlying software, control and cyber-physical systems at an appropriate abstract level. For example, we employed higher-order logic to formalize various control strategies of a platoon due to the involvement of multivariate calculus (i.e, complex frequency domain and Laplace transform). Moreover, we formally verified some of the most important stability constraints for arbitrary platoon parameters. Indeed, this is one of the main strengths of the interactive proof assistants as compared to the simulation based analysis where verification holds only for the applied test cases and thus cannot be considered as complete. However, the verification of important properties of given system in a proof assistant does not guarantee that the system will behave as expected during the runtime operation. Indeed, the verified results in a proof assistant provide a confidence that the system will behave correctly only when the corresponding conditions are met at all times during the life-time of a system. Actually this falls under the scope of runtime verification approaches, which are light-weight formal methods to monitor the correctness of a given system with respect to a formal specification at runtime.

We demonstrate here the utilization of verified stability theorems for various control strategies to construct monitors, which are capable of detecting the violation on a given execution of the system. Consider that the behavior of a platoon controller at each time instant (called an event) is characterized by the tuple \texttt{\textsf{platoon}} and frequency \texttt{\textsf{w}}, i.e., \texttt{\textsf{event = ((x,n,m,k,c,h,ch,vd,h0,ca,cd),w)}}. Thus, an execution of the platoon controller consists of the sequence of events and we model it as an \texttt{\textsf{event list}} in \holl. Next, we consider the autonomous unidirectional controller with constant spacing, for which the stability of the platoon is ensured if the following two conditions are met for every event in the controller execution. 1) \texttt{\textsf{$P_1$ : valid\_platoon \ (x,n,m,k,c,h,ch,vd,h0,ca,cd)}} and 2) \texttt{\textsf{$P_2$ : 0 $<$ $\omega$ $\wedge$  $\mathtt{\mathsf{\dfrac{2 k}{m} \ < \ {\omega}^2}}$}}. In terms of temporal logic, a formal requirement to ensure the platoon stability is $\square P_1 \wedge \square P_2$  where $\square$ represents \emph{Globally} ($G$) or an \emph{Always} operator in the linear temporal logic (LTL). We can model this monitor in \holl~as \textsf{(ALL $P_1$ execution) $\wedge$ (ALL $P_2$ execution)} where  \textsf{ALL} is a \holl~function, which ensures the satisfaction of a predicate on each element of the list. Moreover, we developed a tactic \texttt{\textsf{MONITOR\_TAC}}, which automatically verifies that both the predicates $P_1$ and $P_2$ holds for a given platoon controller execution as a list of events. We tested the efficiency of the \texttt{\textsf{MONITOR\_TAC}} on randomly generated executions and it can check the validity in a reasonable time. For example, on average \texttt{\textsf{MONITOR\_TAC}} returns the truth (T) in $3$ seconds on an execution of $1000$ unique events.

The main purpose of the above illustration was to show that the efforts spent during the formalization within an interactive proof assistant can be complemented by the development of the monitors to ensure the correctness of the system operation at runtime, and thus closing the loop from abstract verification to the real-time monitoring on the concrete system. Our illustration only describes the off-line monitoring where the platoon controller execution is given as a logged data. However, the same monitor can be used for the online monitoring by translating the monitor as a post-condition in the actual system implementation or by generating the monitor using well-known LTL3~\cite{Bauer:2011:RVL:2000799.2000800} or the rewriting-based monitoring approaches~\cite{989799}.

We believe that the stability monitoring can be used for the already available platoon controllers by inspecting the logged traces and by adding monitors in the early controller prototypes for quickly evaluating the correctness of the underlying algorithms. Thus, engineers working on the design and development of the platoon controllers can use the proposed framework without any prior knowledge of theorem proving and gain formally analyzed insights about the given platoon control system.


\vspace*{-0.2cm}

\section{Conclusion and Future Work} \label{SEC:conc}

\vspace*{-0.2cm}

This paper provides a framework for analyzing the platoon control strategies using both the static and dynamic verification approaches. It mainly presents the formal modeling of the platoon controller and its stability using higher-order logic. Next, the proposed formalization is used for formally verifying various platoon control strategies and their stability within the sound core of the \holl~proof assistant. Finally, the formally verified stability theorems are used to develop the runtime monitors for each of the controllers that are used for detecting the violation of any stability constraints.


In future, we plan to formally analyze the platoon considering different connected vehicles (having different masses). We can also incorporate the stability in lateral direction and their physical constraints in our framework for the platoon stability. The other direction is to formally analyze the platoon of connected vehicles, where some of the vehicles act in a malicious manner by changing the control gain and thus the properties of the controllers. Such scenarios can compromise the safety of other vehicles on the highways and result in destabilizing the traffic flow~\cite{dunn2015attacker}. Finally, it is interesting to consider two-dimensional platoons, which can be analyzed by combining our current framework and formalization of the $z$-Transform~\cite{siddique2014formalization}, which is already available in the \holl~proof assistant.

\vspace*{-0.2cm}


\vspace*{-0.3cm}

\bibliographystyle{unsrt}
\bibliography{bibliotex}


\end{document}